\documentclass[aps,prb,twocolumn,showpacs]{revtex4}
\usepackage{graphicx}
\usepackage{graphics}
\usepackage{epstopdf} 

\usepackage[english]{babel}
\usepackage[latin1]{inputenc} 
\usepackage{array}
\usepackage{color}
\usepackage{latexsym}
\usepackage{amsmath,amsxtra,amssymb,amstext,amsfonts}
\usepackage{dsfont}

\begin{document}
\title{Simple analysis of off-axis solenoid fields using the scalar magnetostatic potential: application to a Zeeman-slower for cold atoms}
\author{S\'{e}rgio R. Muniz}
\altaffiliation[Present address: ]{Instituto de F\'{i}sica de S\~{a}o Carlos,
Universidade de S\~{a}o Paulo, S\~{a}o Carlos, SP 13560-970,
Brazil}
\email{srmuniz@ifsc.usp.br}
\affiliation{Joint Quantum Institute, National Institute of Standards and Technology,
and University of Maryland, Gaithersburg, Maryland, 20899, USA}

\author{M. Bhattacharya}
\affiliation{School of Physics and Astronomy,
Rochester Institute of Technology,
84 Lomb Memorial Drive,
Rochester,
NY 14623}

\author{Vanderlei S. Bagnato}
\affiliation{Instituto de F\'{i}sica de S\~{a}o Carlos,
Universidade de S\~{a}o Paulo, S\~{a}o Carlos, SP 13560-970,
Brazil}

\begin{abstract}
In a region free of currents, magnetostatics can be described by the Laplace equation of a scalar magnetic potential, and one can apply the same methods commonly used in electrostatics. Here we show how to calculate the general vector field inside a real (finite) solenoid, using only the magnitude of the field along the symmetry axis. Our method does not require integration or knowledge of the current distribution, and is presented through practical examples, including a non-uniform finite solenoid used to produce cold atomic beams via laser cooling. These examples allow educators to discuss the non-trivial calculation of fields off-axis using concepts familiar to most students, while offering the opportunity to introduce important advancements of current modern research.
\end{abstract}
\date{\today}
\maketitle

\section{Introduction}

Magnetic fields produced by solenoids and axially symmetric coils are ubiquitous, and the ability to calculate them is an integral part of training in physics. Time constraints, however, tend to focus the attention of most introductory electromagnetism (EM) courses to the analytical solution of only a few highly symmetrical cases, such as the field along the axis of a circular coil or inside an infinite solenoid\cite{griffiths1999,Jackson98,Marion95,dasgupta1983,labinac2006}. Nevertheless, many applications require at least an estimate of  the full vector field in regions away from the axis \cite{jackson1999,conway2001}, which involve mathematical tools often not discussed at the introductory level. On the other hand, most EM courses already dedicate a fair amount of time teaching students to identify and solve electrostatic problems using the Laplace equation. In some cases the same methods can be applied to magnetostatic problems, sometimes leading to useful insights.

Sadly, most students do not usually appreciate the similarities between the two classes of problems \cite{lahart2003} due to a limited exposure to practical examples involving the magnetostatic potential. We feel that this ability is useful\cite{dasgupta1985}, particularly because scalar potentials are generally more intuitive and easier to visualize. Besides, a unified treatment could be pedagogically relevant in generalizing the discussion of the multipole expansions\cite{bronzan1971,gray1978,gray1979}. Therefore, the primary goal here is to present a couple of pedagogical examples illustrating the application of the magnetostatic potential method to real solenoids.

In addition, these examples also offer the opportunity to discuss in the classroom axisymmetric fields evaluated off-axis, without the need to introduce the formalism of elliptic integrals. Although other methods for finding off-axis magnetic fields have been mentioned earlier in the literature \cite{jackson1999,conway2001}, to our knowledge this has not been presented from such a simple and intuitive viewpoint.

Moreover, as further motivation, we have chosen an example that brings a real and practical application from the cutting edge of research into the classroom: a non-uniform solenoid used in many research laboratories to produce beams of slow (cold) atoms. This solenoid, called a Zeeman-slower\cite{metcalf1982,MetcalfBook,truscott2004}, is used in conjunction with appropriately prepared laser beams to slow down and cool neutral atoms, from hundreds of Kelvin to milliKelvin temperatures, by combining the action of radiation pressure with the Zeeman effect. This device is one of the staple developments in the area of laser cooling\cite{4,6}, and one of the enabling technologies leading to the 1997 Nobel prize in Physics\cite{PhysNobel}. The techniques for laser cooling and trapping of atoms have produced many dramatic advancements in our understanding of quantum physics\cite{5}, including the achievement of Bose-Einstein condensation\cite{7}, which was recognized with another Nobel prize\cite{PhysNobel} in 2001. In both cases, magnetic fields have been an important part of experimental design and data interpretation. Educators can astutely use the solenoid discussed here, as well as the references herein, to introduce and discuss some of these modern developments in quantum physics, making the subject even more interesting to students.

\section{Reviewing some basic concepts}
\label{sec:Basic}
We begin here by recalling the fundamental equation of magnetostatics : $\overrightarrow{\nabla}\times\overrightarrow{H}=\overrightarrow{J}$, where $\overrightarrow{H}$ is the magnetic field and $\overrightarrow{J}$ the current density. Typically $\overrightarrow{H}$ is related to the magnetic induction field $\overrightarrow{B}$ by some constitutive relation expressing the properties of a particular material. For linear and isotropic materials, with a magnetic permeability $\mu$, $\overrightarrow{B}=\mu\overrightarrow{H}$ and in a current-free region $\overrightarrow{\nabla} \times \overrightarrow{B}=0$, implying that $\overrightarrow{B}=-\overrightarrow{\nabla}\phi_{M}$. Since Maxwell's equations also state that $\overrightarrow{\nabla} \cdot \overrightarrow{B}=0$, this results in $\nabla^{2}\phi_{M}=0$, which is Laplace's equation for the magnetic potential $\phi_{M}$, in any current-free region.

Although Laplace's equation is only typically valid in a region free of charges or currents, these are allowed to exist on or outside a surface $S$ surrounding that region. The solutions of Laplace's equation present three important properties: superposition, smoothness and uniqueness. The property of superposition results from the fact that Laplace's equation is a linear equation. Smoothness implies that no solution in a region $V$ of space, bounded by a surface $S$, can present either a maximum or a minimum within $V$ (extreme values can occur only at the surface $S$). The third property is the one most relevant to us here, as it states \cite{Marion95} that if one finds a solution $\phi_M$, in a region of space consistent with the prescribed boundary conditions, that solution is unique up to an additive constant. Therefore, it does not matter what particular method is used to find the solution. Once an appropriate solution is found, it is uniquely valid.

However, despite the obvious similarities between the electrostatic  and magnetostatic potentials, there are indeed reasons why the analogy can only be taken so far\cite{bronzan1971,lahart2003}, and is not widely explored further in textbooks. The first one arises whenever $\overrightarrow{J}\neq0$, in which case it is not trivial to write a relation between $\phi_M$ and $\overrightarrow{J}$. The second complication occurs due to the fact that the scalar potential is generally a multiply valued function, requiring a prescription specifying where it can be used. However, as it has been shown by Bronzan\cite{bronzan1971}, these complications can be overcome, permitting one to exploit the advantages of the  concept of a scalar magnetic potential.

\section{The magnetic field of a finite uniform solenoid}
\label{sec:Finite}

\begin{figure} [l 5 h!]
\begin{center}
\includegraphics [width = 0.95 \columnwidth]{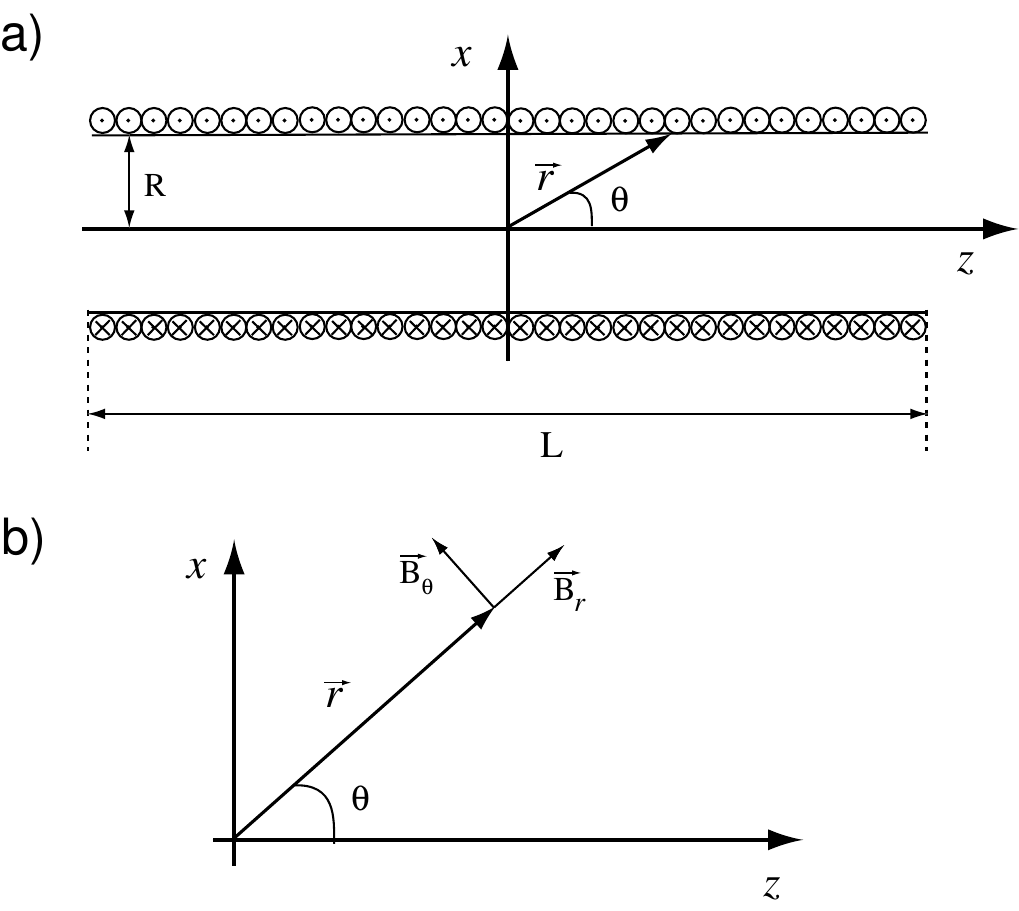}
\end{center}
\vspace{-6mm}
\caption{\label{Fig1} Schematic representation of a finite and homogenous solenoid.
The relevant coordinates are presented in (a), while their details are shown in (b).
The crosses (dots) represent current flowing into (out of) the plane of the page.}
\vspace{-3mm}
\end{figure}

We start by considering the field in the interior of a finite uniform solenoid carrying a current $I$, as illustrated in Fig.~\ref{Fig1}. For generality and convenience, we describe the problem using spherical coordinates. In this geometry it is easy to note that due to the axial symmetry of the problem the field $B(z)$ depends only on $z=r\cos \theta$. The magnetostatic potential can be found using for a boundary condition the magnitude of the field along $z$, which is readily available through simple summation formulas over the approximately circular coils forming the solenoid.

In spherical coordinates, the solution of the axisymmetric  scalar potential $\phi_{M}$  can be written in the
form:
\begin{equation}
\phi_{M}(r,\theta)=\sum_{\ell=0}^{\infty}\left(a_{\ell}r^{\ell}+
\frac{b_{\ell}}{r^{\ell+1}}\right)P_{\ell}(\cos \theta),
\label{Eq.1}
\end{equation}
where $a_{\ell}$ and $b_{\ell}$ are coefficients to be determined
and the $P_{\ell}$ represents a Legendre polynomial of order $\ell$.

Because we are mainly interested in the values of the field inside the solenoid, we set $b_{\ell}=0$ to avoid the singularity at $r=0$. As a result, the potential takes the simpler form:
\begin{equation}
\phi_{M}(r,\theta)=a_{0}+a_{1}rP_{1}(\cos \theta)
+a_{2}r^{2}P_{2}(\cos \theta)+ \ldots .
\label{Eq.2}
\end{equation}

For points along the $z$-axis, we have $\cos \theta =1$ and Eq.~(\ref{Eq.2}) becomes
\begin{equation}
\phi_{M}(z)=a_{0}+a_{1}z+a_{2}z^{2}+a_{3}z^{3}+ \ldots .
\label{Eq.3}
\end{equation}

Incidentally, we can in general also expand the scalar potential in a Taylor series around some point $z_0$,
\begin{equation}
\phi_{M}(z)=\phi_{M}(z_0)+(z-z_{0})\frac{\partial\phi_{M}}{\partial z}\mid_{z_0} +\frac{(z-z_{0})^{2}}{2!}\frac{\partial^{2}\phi_{M}}{\partial z^{2}}\mid_{z_0}+\ldots ,
\label{Eq.4}
\end{equation}
and comparing it to Eq.~(\ref{Eq.3}), for $z_0=0$, we obtain the coefficients $a_{\ell}$ in terms of the series expansion
\begin{equation}
a_{0}=\phi_{M}(0);\text{  }\ldots;\text{  } a_{\ell}=\frac{1}{\ell!}\frac{\partial^{\ell}\phi_{M}(z)}{\partial z^{\ell}
}\mid_{z=0}.
\label{Eq.5}
\end{equation}
In this way, the full scalar potential becomes analytically determinable, allowing us to evaluate $\overrightarrow{B}=-\overrightarrow{\nabla}\phi_{M}$ at any point in space where equations ~(\ref{Eq.2}) and ~(\ref{Eq.4}) appliy.

As a first example, let us now consider the case of a finite solenoid of length $L$ and radius $R$, carrying a uniform current $I$. If the solenoid has $N$ turns per unit length, the magnetic field along the $z$-axis can be easily calculated by integrating the expression for the axial field of a circular current loop \cite{griffiths1999}, resulting in:
\begin{equation}
B(z)=\alpha \left[\frac{z_{+}}{\sqrt{R^{2}+z_{+}
^{2}}}-\frac{z_{-}}{\sqrt{R^{2}+z_{-}^{2}}}\right],
\label{Eq.6}
\end{equation}
where $z_{\pm}=z \pm \frac{L}{2}$, and $\alpha= \mu_0 NI/4\pi$ in SI  units.
Now, since
\begin{equation}
B_z(z)=-\frac{\partial\phi_{M}}{\partial z},
\label{Eq.7}
\end{equation}
we can write
\begin{equation}
\phi_{M}(z)=-\int B\left(  z^{\prime}\right)  dz^{\prime}.
\label{Eq.8}
\end{equation}
Using Eq.~(\ref{Eq.6}) in Eq.~(\ref{Eq.8}) we obtain the general form of the potential for the finite solenoid along the axis:
\begin{equation}
\phi_{M}(z)=-\alpha\left[ \left(  R^{2}+z_{+} ^{2}\right)
^{\frac{1}{2}}-\left(  R^{2}+z_{-}^{2}\right) ^{\frac{1}{2}}\right].
\label{Eq.9}
\end{equation}

Expanding Eq.~(\ref{Eq.9}) around $z=0$, as in Eq.~(\ref{Eq.4}), we get the various coefficients for $\phi_{M}(z)$. Using these coefficients and introducing the expression for the Legendre polynomials $P_{\ell}(\cos \theta)$, while keeping terms up to third order, we finally get:
\begin{align}
\phi_{M}(r,\theta) & =-\frac{\alpha L r \cos \theta}
{\left(\frac{L^2}{4}+R^2\right)^{1/2}} +
   \left(\frac{5}{2} \cos^3 \theta - \frac{3}{2} \cos \theta \right) \\ & \times
   \left[\frac{L \alpha}{2\left(\frac{L^2}{4}+R^2\right)^{3/2}}
   -\frac{L^3 \alpha }{8 \left(\frac{L^2}{4}+R^2\right)^{5/2}}\right]r^3 +\dots ,\ \ \nonumber
\label{Eq.10}
\end{align}

From the last equation, one can calculate the components $\ B_{\theta}%
\ $and $B_{r}$ of the magnetic field by simply taking the gradient of the potential:
\begin{align}
B_{r}  &  = -\frac{\partial\phi_{M}(r,\theta)}{\partial r} \\ \nonumber
& = \frac{\alpha L \cos \theta}{\left(\frac{L^2}{4}+R^2 \right)^{1/2}}-
\frac{3}{2} r^2 \left(5 \cos^3 \theta -3 \cos \theta \right) \\
& \times \left[\frac{L \alpha}{2 \left(\frac{L^2}{4}+R^2\right)^{3/2}} -\frac{L^3 \alpha }{8 \left(\frac{L^2}{4}+R^2\right)^{5/2}}\right]+\dots ,\nonumber
\label{Eq.11}
\end{align}
and
\begin{align}
B_{\theta} & = -\frac{1}{r}\frac{\partial\phi_{M}(r,\theta)}{\partial\theta} \\
\nonumber & = -\frac{L \alpha  \sin \theta
}{\left(\frac{L^2}{4}+R^2\right)^{1/2}} - \left(\frac{3}{2} \sin \theta-
\frac{15}{2}
\cos ^2 \theta \sin \theta \right) \\
& \times\left[\frac{L \alpha } {2\left(\frac{L^2}{4}+R^2\right)^{3/2}} -\frac{L^3 \alpha }{8 \left(\frac{L^2}{4}+R^2\right)^{5/2}}\right]r^2 +\dots \nonumber
\label{Eq.12}
\end{align}

It is interesting to note that these are approximate analytical results for the magnetic field inside the solenoid, provided it is within the radius of convergence of the power series and away from the current paths (wires), with their precision limited by the number of terms included in the expansion.

One can test these results by comparing the expressions (\ref{Eq.10}) and (\ref{Eq.11}) with those presented in Chapter 5 of ref. \cite{Jackson98}, where a different method was used to evaluate the field components. In particular, we will show that if one keeps only the first order in the expansion, the result simplifies to the approximate solution of problem 5.2 (b) in the 2nd ed. of ref. \cite{Jackson98}. For that we recall the relations
\begin{align}
B_{\rho} &= B_{r}\sin\theta+B_{\theta}\cos\theta,\\
B_{z} &= B_{r}\cos\theta-B_{\theta}\sin\theta, \nonumber \ \
\label{Eq.13}\
\end{align}
from which we obtain, up to third order,
\begin{equation}
B_{\rho} \simeq \frac{3\alpha LR^{2}r^{2}}{2[R^{2}+(L/2)^{2}]^{5/2}}\sin\theta
\cos\theta.
\label{Eq.14}
\end{equation}

Finally, using $r\sin\theta=\rho$ and $r\cos\theta=z$, in the limit $R/L\ll1$, Eq.~(\ref{Eq.14}) yields
\begin{equation}
B_{\rho}(z,\rho)\simeq\frac{96\pi NIR^2}{c}\frac{\rho z}{L^{4}},
\label{Eq.15}
\end{equation}
which is expressed here in CGS (Gaussian) units, with $\alpha=2\pi NI/c$, to facilitate a direct comparison with the result presented in the second edition of reference \cite{Jackson98}.

\begin{figure} [r t !]
\begin{center}
\includegraphics [width = 0.95 \columnwidth]{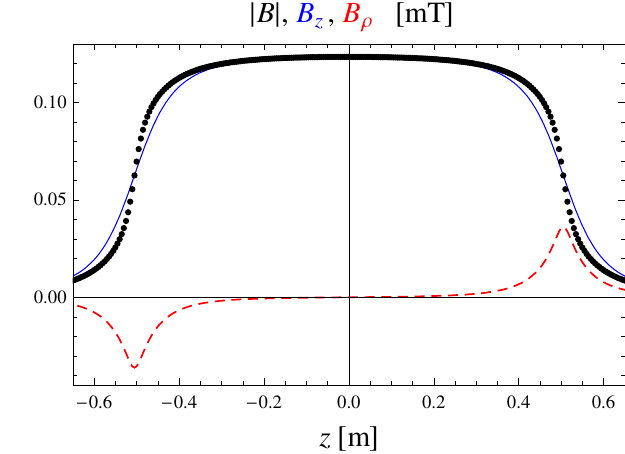}
\vspace{-.2in}
\end{center}
\caption{\label{Fig2} Numerical calculation of the exact solution (using elliptic integrals) for the field of a finite solenoid of length $L=1 \mbox{ m}$, radius $R=10 \mbox{ cm}$, number of windings $N=100$ and carrying a current $I=1 \mbox{ A}$. The points (black) show the magnitude $|B|$ along the axis ($\rho=0$), whereas the solid (blue) line represents $B_z$, and dashed (red) line $B_{\rho}$ at $\rho=8 \mbox{ cm}$. }
\end{figure}

To show a practical application of the method, we now compare our results to a realistic numerical calculation of a finite solenoid. Figure \ref{Fig2} shows the numerical results for the magnitude of the magnetic field and its components. The calculation assumes that the solenoid is composed of a series of circular coils, and performs a direct summation over the exact analytical expression, based on elliptic integrals, for each individual coil. The numerical result were verified to accurately represent the field of a physical solenoid, through measurements with a Hall probe along the axis. This was expected since the error introduced by approximating the actual helical winding by a sequence of circular coils is typically negligible at this scale.
The relevant physical parameters are given in the captions of Fig.\ref{Fig2}.

\begin{figure} [h!]
\begin{center}
\includegraphics [width = 1 \columnwidth]{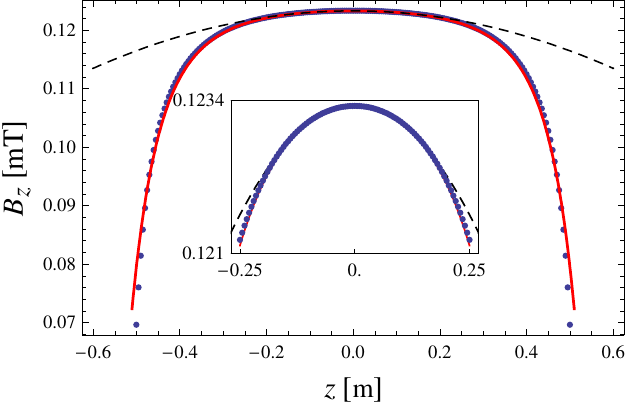}

\vspace{+.3in}
\includegraphics [width = 1 \columnwidth]{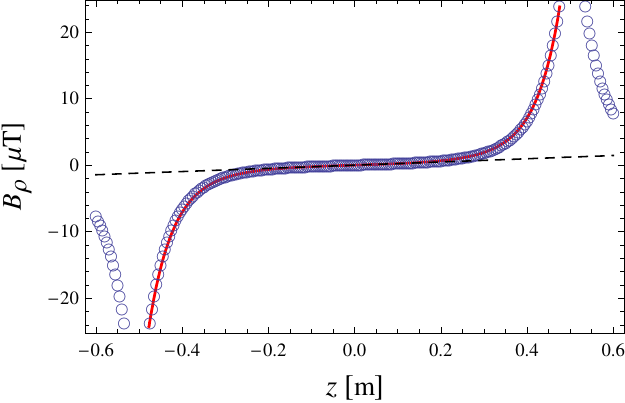}

\end{center}
\vspace{-.1in}
\caption{\label{Fig3} Comparison of numerical (points) and
analytical (lines) results for the uniform solenoid. In (a)  the axial and in (b) the transverse field profiles have been shown. All the results were evaluated off-axis, at  $\rho=8 \mbox{ cm}$, and the analytical results are shown for expansions up to third (dashed-black) and fifteenth (solid-red) orders. The inset in (a) shows the good agreement obtained near z=0.}
\end{figure}

Further, in Fig.~\ref{Fig3}, we compare the analytical results obtained by keeping the first eight terms in the expansion (corresponding to the fifteenth order in $r$) against the numerical calculations shown in Fig.~\ref{Fig2}. In addition, Fig.~\ref{Fig3} also shows the good partial agreement obtained using the third order approximation, extending to distances up to about half the size of the solenoid. Note that, for a real finite system, the disagreement increases rapidly after some point. That can be improved significantly by including higher order terms, allowing for a much better approximation near the edges, as shown in Fig.~\ref{Fig3}. However, due to the simplifications made, the approximate analytical result does not contain all the physics of the problem. For instance, it does not accurately describe the field outside the solenoid, and it will most likely fail outside the radius of convergence of the power series expansion. Nevertheless, the magnetostatic potential method still provides a reasonable representation of the internal fields up to the very end of the solenoid.

\section{The Zeeman-slower: an inhomogeneous finite solenoid}
\label{sec:Zeeman}

\begin{figure} [tl]
\vspace{-.1in}
\begin{center}
\includegraphics [width = 0.8    \columnwidth]{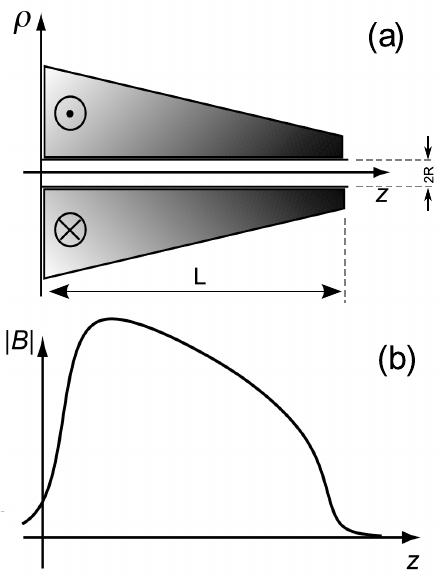}
\vspace{-.35in}
\end{center}
\caption{\label{Fig4} (a) Sketch of a tapered (triangular shape) solenoid creating an inhomogeneous current distribution to produce the appropriate axial field profile (b) for a Zeeman-slower.}
\end{figure}

Now we will consider another interesting and very practical problem, familiar to many atomic physics laboratories, namely the design of a solenoid capable of producing an axial field with a parabolic profile, as in
\begin{equation}
\label{Eq.16}
B(z)=B_{b}+B_{0}\sqrt{1-\beta z},
\end{equation}
where $B_{b}$, $B_{0}$ and $\beta$ are constants. Such a field is suitable for slowing atomic beams using laser light \cite{metcalf1982}. The field of Eq.~(\ref{Eq.16}) causes a spatially varying Zeeman effect that compensates for the changing Doppler shift of the moving atoms, thus keeping them in resonance with the light as they decelerate along the beam path. This technique is called Zeeman slowing\cite{metcalf1982}, and the parabolic shape is chosen to keep the radiation pressure constant, typically at a rate of $\sim10^6$  $m/s^2$, throughout the Zeeman solenoid\cite{MetcalfBook} shown in Figure~\ref{Fig4}.

In general, the atomic beam encompasses a certain solid angle as it traverses the solenoid and most atoms follow trajectories which do not lie exactly on the axis. Since the resonance condition with the laser depends on both the magnitude (via the detuning) and direction (via the polarization) of the magnetic field, the knowledge of the off-axis field is important in understanding how light interacts with atoms at different points inside the solenoid.

The magnetic potential along the z-axis, in this case, takes the form
\begin{equation}
\phi_{M}(z)=-\int B(z')dz'=-B_{b}z+\frac{2}{3}\frac{B_{0}}{\beta}(1-\beta
z)^{\frac{3}{2}},
\label{Eq.17}
\end{equation}
where the constant of integration has been suppressed.

Following the same steps in section ~\ref{sec:Finite}, and after calculating the derivatives and solving for the coefficients $a_{\ell}$, we obtain the general form of the magnetic potential for the Zeeman solenoid:
\begin{align}
\phi_{M}(r,\theta)  &  =\frac{2}{3}\frac{B_{0}}{\beta}-(B_{b}+B_{0})r\cos \theta \\
&
+\frac{B_{0}}{\sqrt{\pi}}\sum_{n=2}^{\infty}\frac{\beta^{n-1}\Gamma\left(n-\frac{1}{2}\right)}{n!(2n-3)}
r^{n}P_{n}(\cos\theta) \nonumber,
\label{Eq.18}
\end{align}
where\ $\Gamma(n)$ is the gamma function. Now we can calculate the spherical components of the magnetic field,
\begin{align}
B_{r}(r,\theta) & =(B_{b}+B_{0})\cos\theta  \\
&-\frac{B_{0}}{\sqrt{\pi}}
\sum_{n=2}^{\infty}\frac{\beta^{n-1}\Gamma\left(
n-\frac{1}{2}\right)}{(n-1)!(2n-3)} r^{n-1}P_{n}(\cos\theta) \nonumber,
\label{Eq.19}
\end{align}
and
\vspace{-.2in}
\begin{align}
B_{\theta}(r,\theta)& =-(B_{b}+B_{0})\sin \theta   \\
&-\frac{B_0} {\sqrt{\pi} \sin \theta} \sum _{n=2}^{\infty }
\frac{\beta ^{n-1} \Gamma \left(n-\frac{1}{2}\right)}
{(n-1)! (2n-3)}r^{n-1}   \nonumber \\
&\times \left[\cos \theta P_n(\cos \theta)-P_{n-1}(\cos \theta
)\right].    \nonumber
\label{Eq.20}
\end{align}

\begin{figure} [tl]
\begin{center}
\includegraphics [width = 0.9 \columnwidth]{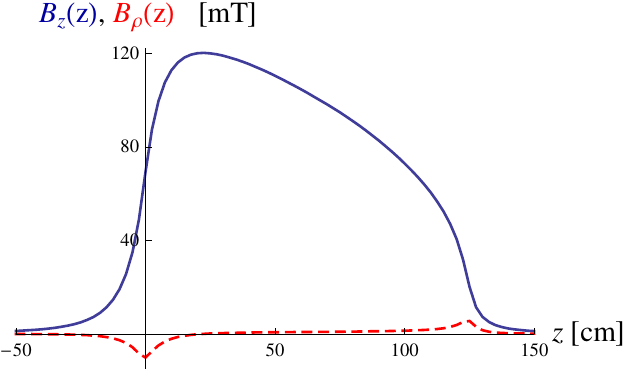}
\vspace{-.2in}
\end{center}
\caption{\label{Fig5} Numerical results representing the experimental field of a real Zeeman solenoid, with $L=125 \mbox{ cm}$, $R=4 \mbox{ cm}$, and $B_{max}\approx120 \mbox{ mT}$.  The solid (blue) line shows $|B|$ along the axis ($\rho=0$), whereas the dashed (red) line represents the transverse field $B_{\rho}$ at $\rho=2.5 \mbox{ cm}$. }
\end{figure}

The transverse and axial components can be easily obtained from Eq.~(13), with the shorthand $\tilde{z}=\frac{\displaystyle z}{\displaystyle \sqrt{z^2+\rho^2}}$:
\begin{align}
B_{\rho}(\rho,z) = \frac{-B_o}{\sqrt{\pi}}& \sum_{n=2}^{\infty}
   \frac{\beta ^{n-1} \Gamma \left(n-\frac{1}{2}\right)}{(n-1)!(2n-3)} \\
&\times \frac{( \sqrt{z^2+\rho^2} )^{n}}{\rho} \left[P_{n}(\tilde{z})-\tilde{z} P_{n-1}(\tilde{z})\right], \nonumber
\label{Eq.21}
\end{align}
and
\vspace{-.1in}
\begin{align}
B_{z}(\rho,z)& =(B_{b}+B_{0}) \\
&-\frac{B_0} {\sqrt{\pi}} \sum _{n=2}^{\infty }
\frac{\beta ^{n-1} \Gamma \left(n-\frac{1}{2}\right)}
{(n-1)! (2n-3)}( \sqrt{z^2+\rho^2} )^{n-1} P_{n-1}(\tilde{z}). \nonumber
\label{Eq.22}
\end{align}

Notice that the transverse component of the magnetic field inside the Zeeman-slower does not depend on $B_{b}$. Also, note that $B_{\rho}=0$ at $\rho=0$ (on-axis), as expected. It can be verified that the on-axis field sums back to the exact expression of Eq.~(\ref{Eq.16}). Although caution may be necessary when evaluating the field for $\rho=0$  ($\theta=0$), the careful use of L'Hospital's rule ensures correct answers.

\begin{figure} 
\begin{center}
\includegraphics [width = 0.95 \columnwidth]{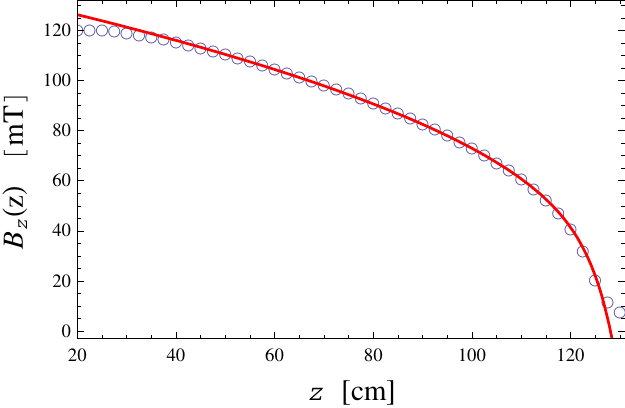}

\includegraphics [width = 0.95 \columnwidth]{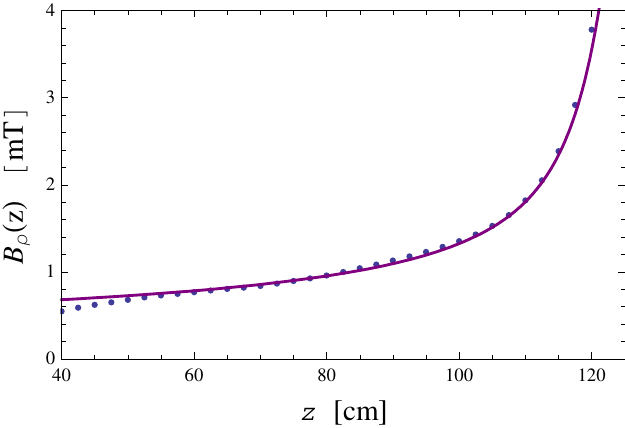}
\end{center}
\vspace{-.2in}
\caption{\label{Fig6} Comparison between numerical (from Figure \ref{Fig5}) and
analytical solutions for the Zeeman solenoid for (a) $B_z$ for $\rho=0$, and (b) $B_{\rho}$ for $\rho=2.5 \mbox{ cm}$. Points (blue)represent the experimental field and solid lines (red, purple) the analytical approximations of Eq.(22) and Eq.(21) respectively.}
\vspace{-.1in}
\end{figure}

Now, to compare these analytical approximations to the experimental field represented in Fig. \ref{Fig5}, we will follow a different approach. The motivation here is to mimic a situation where the current distribution that generates the field may not be known exactly, but the axial field can be measured directly in the laboratory. This could be the case in a real practical application, where imperfections in the winding pattern often are not considered in the ideal model. From the experimental data one can then build a mathematical model, using either a fitting function (if the functional form is known or could be easily guessed), or by using an interpolating function, such as a polynomial, to represent the data in a limited region of space. Here, since the approximate functional form of the axial field is known, we will extract the model parameters by numerically fitting the data in Fig.~\ref{Fig5}, to $B_z(z)$ in Eq.~(\ref{Eq.16}), and substituting them in the expressions (21) and (22). Note that the limitations of this type of modeling may result in some inaccuracies, particularly close to the edges. In any practical situation one may need to explore different approaches to find a  mathematical model accurate enough in the region of interest.

After following these steps to model the data, we show in Fig.~\ref{Fig6} a comparison between our analytical approximations and the numerical result (Fig.~\ref{Fig5}), that represent very accurately the experimental field, as determined by measurements in our laboratory. There is a reasonable agreement between the solid lines, representing equations (21) and (22) and the data points, representing the numerically calculated field. Note that, in contrast to the uniform finite solenoid where the power series was used to approximate the exact solution, here the power series simply approximates our model\cite{8} function. Therefore, increasing the order\cite{9} of the expansion only improves the agreement with the model (fitting) function, which represents the data only over a limited region and does not contain all the information about the fields in the problem. This is clearly visible in Fig.~\ref{Fig6} where good agreement is found only in the range $z\approx$ (0.4 to 1.2) m.

\section{\textbf{Conclusion}}
\label{sec:Discussion}

Using the simple concept of the magnetostatic scalar potential, and only the knowledge of the field along the symmetry axis, we have shown how to determine the vector magnetic field anywhere inside an inhomogeneous finite solenoid, without explicitly integrating (or even knowing) the current distribution.
In cases where the current distribution is known, but the expression for the field off-axis is non-trivial (for instance, given by elliptical integrals), one can still gain some insight by using the method described here. This simple analysis follows from a straightforward analogy with the electrostatic boundary value problem, and can be useful in determining field inhomogeneities in various practical experiments involving solenoids. In the present article we have used an example from contemporary atomic physics experiments to demonstrate the method. However, we believe that a simplified version of this discussion (e.g.: the uniform finite solenoid) could be used in a regular classroom setting, as a practical example of a calculation of off-axis magnetic fields, for undergraduate courses and teaching laboratories in physics and engineering.


\begin{thebibliography}{}
\bibitem{9} For the agreement shown in Fig.~\ref{Fig6}, 70 terms were kept.
\end{thebibliography}
\end{document}